\documentclass[pra,twocolumn,nofootinbib,floatfix,10pt,aps]{revtex4-2}

\usepackage{amsmath}
\usepackage{amssymb}
\usepackage{wasysym}
\usepackage{graphicx}
\usepackage{color,soul}
\usepackage{physics}
\usepackage{siunitx}
\usepackage{dsfont}
\usepackage{float}
\usepackage[hidelinks,colorlinks=true,linkcolor=black,citecolor=black,urlcolor=black]{hyperref}
\usepackage[english]{babel}

\usepackage[normalem]{ulem}

\long\def\symbolfootnote[#1]#2{\begingroup
\def\thefootnote{\fnsymbol{footnote}}\footnotetext[#1]{#2}\endgroup}

\newcommand{\tpbare}{\ensuremath{\tilde{t}_{\perp}}}
\newcommand{\tplbare}{\ensuremath{\tilde{t}_{\parallel}}}
\newcommand{\Jpl}{\ensuremath{J_{\parallel}}} 
\newcommand{\Jp}{\ensuremath{J_{\perp}}}
\newcommand{\JpD}{\ensuremath{J_{\perp}}}
\newcommand{\tp}{\ensuremath{t_{\perp}}}
\newcommand{\tpl}{\ensuremath{t_{\parallel}}}
\newcommand{\gh}{\ensuremath{g_{\mathrm{h}}^{(2)}}}
\newcommand{\gho}{\ensuremath{g_{\mathrm{h}}^{(2)}}}
\newcommand{\gp}{\ensuremath{g_{\mathrm{pair}}^{(2)}}}
\newcommand{\kB}{\ensuremath{k_{\rm B}}} 
\newcommand{\Tc}{\ensuremath{T_{\mathrm{C}}}}
\DeclareSIUnit\Gauss{G}

\begin{document}

\title{Magnetically mediated hole pairing in fermionic ladders of ultracold atoms}

\author{Sarah~Hirthe$^{1,2,\ast}$}
\author{Thomas~Chalopin$^{1,2}$}
\author{Dominik~Bourgund$^{1,2}$}
\author{Petar~Bojovi\'{c}$^{1,2}$}
\author{Annabelle~Bohrdt$^{3,4}$}
\author{Eugene~Demler$^{5}$}
\author{Fabian~Grusdt$^{2,6,7}$}
\author{Immanuel~Bloch$^{1,2,7}$}
\author{Timon~A.~Hilker$^{1,2}$}

\affiliation{$^1$Max-Planck-Institut f\"{u}r Quantenoptik, 85748 Garching, Germany}
\affiliation{$^2$Munich Center for Quantum Science and Technology, 80799 Munich, Germany}
\affiliation{$^3$Department of Physics, Harvard University, Cambridge, MA 02138, USA}
\affiliation{$^4$ITAMP, Harvard-Smithsonian Center for Astrophysics, Cambridge, MA 02138, USA}
\affiliation{$^5$Institute for Theoretical Physics, ETH Zurich, 8093 Zurich, Switzerland}
\affiliation{$^6$Department of Physics and Arnold Sommerfeld Center for Theoretical Physics (ASC), Ludwig-Maximilians-Universit\"{a}t, 80799 Munich, Germany}
\affiliation{$^7$Department of Physics, Ludwig-Maximilians-Universit\"{a}t, 80799 Munich, Germany}

\symbolfootnote[1]{Electronic address: {sarah.hirthe@mpq.mpg.de }}

\begin{abstract}
Pairing of mobile charge carriers in doped antiferromagnets plays a key role in the emergence of unconventional superconductivity~\cite{lee:2006}.
In these strongly correlated materials, the pairing mechanism is often assumed to be mediated by magnetic correlations~\cite{Scalapino1999}, in contrast to phonon-mediated interactions in conventional superconductors~\cite{bardeen:1957}.
A precise understanding of the underlying mechanism in real materials is, however, still lacking, and has been driving experimental and theoretical research for the past 40 years.
Early theoretical studies established the emergence of binding among dopants in ladder systems~\cite{dagotto:1992,Sigrist1994,Troyer1996,Schulz1999,Giamarchi2004}, where idealised theoretical toy models played an instrumental role in the elucidation of pairing, despite repulsive interactions~\cite{Kohn1965}. 
Here, we realise this long-standing theoretical prediction and report on the observation of hole pairing due to magnetic correlations in a quantum gas microscope setting.
By engineering doped antiferromagnetic ladders with mixed-dimensional couplings~\cite{bohrdt:2021} we suppress Pauli blocking of holes at short length scales.
This results in a drastic increase in binding energy and decrease in pair size, enabling us to observe pairs of holes predominantly occupying the same rung of the ladder. We find a hole-hole binding energy on the order of the superexchange energy, and, upon increased doping, we observe spatial structures in the pair distribution, indicating repulsion between bound hole pairs.
By engineering a configuration in which binding is strongly enhanced, we delineate a novel strategy to increase the critical temperature for superconductivity.
\end{abstract}
\maketitle

\begin{figure*}[t]
\centering
\includegraphics{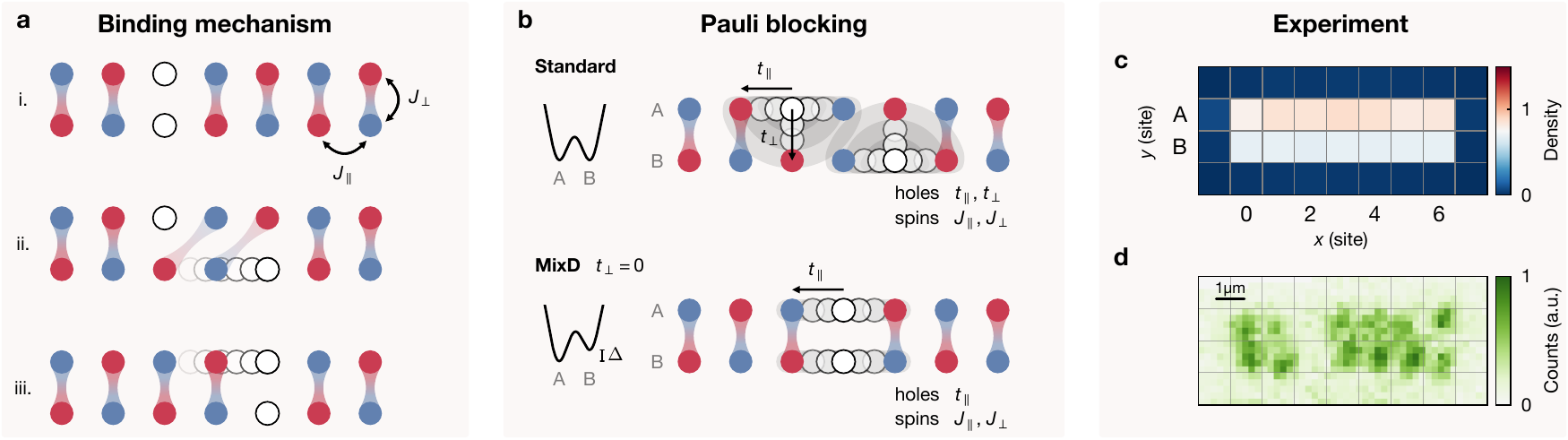}
\caption{
\textbf{Hole-pairing in mixed-dimensional ladders.}
\textbf{a}, Binding mechanism in the $t-J$ ladders.
Depicted are ladder systems with spin exchange $\Jp\gg\Jpl$ that form strong singlet bonds along the rungs.
When a single hole from (i) moves through the system, as illustrated in (ii), it breaks the spin order by displacing the singlet bonds.
(iii) The magnetic energy cost can be avoided if the second hole restores the spin order by moving together with the first hole.
\textbf{b}, Pauli blocking of holes.
Due to their fermionic nature, holes repel each other along all directions according to the tunnelling amplitudes $\tp$, $\tpl$.
Close-distance hole pairs are thus energetically unfavourable.
In mixD systems a potential offset $\Delta$ between the two legs suppresses tunnelling $\tp$ and Pauli repulsion only occurs along the legs.
Holes on the same rung can thus benefit from the binding mechanism, forming tightly bound pairs with large binding energy.
\textbf{c}, Average density of the mixD $L=7$ ladder system with $\Delta \approx U/2$.
\textbf{d}, Single experimental shot with two holes on the same rung, exemplifying the bunching of holes in the mixD system.
}
\label{fig:fig1}
\end{figure*}

Unconventional superconductivity in materials like heavy fermion systems~\cite{White2005}, iron pnictides~\cite{Wen2011}, layered organic materials~\cite{Wosnitza2012}, cuprate superconductors~\cite{keimer:2015}, and twisted bilayer graphene~\cite{Cao2018}, arises in the vicinity of magnetically ordered states.
A common mechanism consisting of dopant pairing mediated by magnetic fluctuations is thus believed to be at the heart of these superconducting states~\cite{Scalapino1999}, but a detailed understanding of the underlying physics remains a key problem in quantum many-body physics.

A promising tool to explore strongly correlated quantum systems is analogue quantum simulation using ultracold atoms~\cite{gross:2017}. 
Recent experimental progress investigating doped antiferromagnets has been made using single-site resolved fermionic quantum gas microscopes~\cite{chiu:2019,brown:2019,koepsell:2019,ji:2021,koepsell:2021, bohrdt:2021a}, which provide microscopic real-space correlations complementary to the spectroscopic and transport measurements performed in solids. They typically simulate the Fermi-Hubbard model, consisting of itinerant spin-$1/2$ fermions within a single band of a periodic lattice.

Even though the two-dimensional Fermi-Hubbard model displays many characteristics also found in the high-$\Tc$ superconducting cuprates~\cite{lee:2006}, the existence of pairing and superconductivity in this model remains a subject of debate~\cite{Qin2020,Arrigoni2004}.
While the magnetic degree of freedom favours the pairing of holes, the bound state is frustrated due to Pauli blocking causing a kinetic energy cost for two holes at close distance.
This intricate interplay of magnetic and kinetic energy not only complicates theoretical investigations but also also leads to small binding energies~\cite{Leung2002, blomquist:2021} rendering pairing challenging to observe at temperatures of state-of-the-art quantum simulators~\cite{mazurenko:2017a}.

In order to shed light on the pairing mechanism in doped Mott insulators, several theoretical studies considered doped Fermi Hubbard and $t-J$ ladders~\cite{dagotto:1992,Sigrist1994,Troyer1996,Chernyshev1998,Giamarchi2004}, where accurate numerical solutions can be obtained using DMRG~\cite{White1992}.
In solid-state experiments, ladder materials have also been shown to display superconductivity~\cite{Uehara1996,Nagata1998, Dagotto1999}.
A paradigmatic case for theoretical investigation, that exhibits large binding, is the regime where inter-chain magnetic exchange is larger than single-particle interchain hopping~\cite{Schulz1999}. These parameters could, however, not be justified microscopically for condensed matter systems and are unphysical within the framework of a pure Fermi-Hubbard system (see SI). 
The key motivation for our work was to provide an experimental realisation of this system that has been considered only a theoretical abstraction.
We achieve this by extending Fermi-Hubbard ladders at large interactions with a potential offset, effectively realising a  mixed-dimensional (mixD) system~\cite{bohrdt:2021}. 

The essential physics of our experiment is captured (see SI) by the $t-J$ Hamiltonian
\begin{equation}
\begin{aligned}
\hat{\mathcal{H}} = &  \sum_{\langle i , j\rangle,\sigma}\hat{\mathcal{P}} \left( - t_{ij}\, \hat{c}^{\dagger}_{i,\sigma}\hat{c}^{\vphantom{\dagger}}_{j,\sigma} + \text{h.c.}\right)\hat{\mathcal{P}} \,+ \\
& + \sum_{\langle i , j\rangle} J_{ij} \left( \hat{\mathbf{S}}_{i} \cdot \hat{\mathbf{S}}_{j} - \frac{\hat{n}_i\hat{n}_j}{4}\right),
\end{aligned}
\label{eqn:tJ}
\end{equation}
where $\hat{\mathcal{P}}$ projects to the subspace without double occupancies; the hopping energy is $t_{ij} = t_\parallel$ ($t_\perp$) and the superexchange energy is $J_{ij} = J_\parallel$ ($\Jp$) for nearest-neighbour sites $i,j$ on the same leg (rung).
$\hat{c}^{\dagger}_{i,\sigma}$ ($\hat{c}_{i,\sigma}$) denotes the creation (annihilation) operator of a fermion on site $i$ with spin $\sigma = \uparrow,\downarrow$; $\hat{\mathbf{S}}_{i}$ and $\hat{n}_i$ are the on-site spin and density operators.
The mixD system is described by Eq. (\ref{eqn:tJ}) for $\tp = 0$, which we realise by suppressing tunnelling, but not spin exchange, using a potential offset along the rungs.
We consider both the pure Fermi-Hubbard system (in the following denoted as standard ladder) and the mixD system in the regime of strong spin coupling along the rung $\Jp \gg \Jpl$ with a high singlet fraction along the rungs.

The pairing of holes in a model exhibiting only repulsive interactions~\cite{Chakravarty2001,Kantian2019}, arises from the singlet spin gap, which has a conceptual connection to the RVB mechanism~\cite{anderson:1987}. 
Binding can be understood from the competition of hole delocalisation and spin order along the rungs, as illustrated in Fig.~\ref{fig:fig1}a. 
If a hole moves through the system, it displaces these spins creating an energetically unfavourable magnetic configuration.
Therefore a single hole becomes dressed by a cloud of disturbed correlations and forms a polaron  with reduced mobility~\cite{koepsell:2019,ji:2021}.
However, if a second hole moves along with the first hole, it can restore the order in the spin sector, causing the two holes to form a highly mobile bound pair.
Hole pairing extends to the limit of uncoupled rungs ($\tpl \ll \Jp$) because two holes on the same rung maximise the number of singlets that can form in the ladder.

However, this process only dominates in the mixD case, whereas in the standard ladders $\tp$ is the dominating energy scale. In the latter, binding competes with the repulsion of holes due to Pauli blocking, rendering tight pairs energetically unfavourable (see \hyperref[fig:fig1]{Fig.\,1b}).
Pairing between holes can still occur, but only with a small energy advantage $ E_\mathrm{b}\ll \Jp$ and at large pair sizes~\cite{white:1997a}.
By realising mixD ladders, we strongly suppress the Pauli repulsion between holes along the rung (see Fig.\,\ref{fig:fig1}b), thus engineering a system with strong hole attraction.
This binding mechanism is protected by the spin gap and thus persists up to  high temperatures on the order $\Jp$.

We realise ladders of length $L = 7$ in our Fermi-gas microscope with independently tunable optical lattices and single-site resolved optical potential shaping.
The mixD $t-J$ system is derived as an effective model from fermionic atoms in a two-leg ladder-shaped lattice potential described by the Hubbard parameters $U$ (on-site interaction), $\tplbare,\tpbare$ (tunnelling amplitudes) and with a potential offset $\Delta$ between the two legs.
For large $U/\tplbare,U/\tpbare$, it is effectively described by the $t-J$ model and its distinct parameters as in Eq. (\ref{eqn:tJ}). 
For $U>\Delta>\tpbare$, tunnelling along the rungs is suppressed to $\tp = 0$, while tunnelling along the leg is unaffected, $\tpl=\tplbare$, and spin exchange is given by $\Jp= 2\tpbare\hspace{-0.2em}^2/(U+\Delta)+2\tpbare\hspace{-0.2em}^2/(U-\Delta)$~\cite{duan:2003, trotzky:2008} and $\Jpl = 4\tplbare \hspace{-0.1em} ^2/U$. 

Our experiment begins by preparing a balanced mixture of the lowest two hyperfine states of $^6$Li, which we load into an engineered ladder geometry similar to our previous work~\cite{sompet:2021}.
In order to equally distribute doped holes over both legs of the final system, we first load atoms into uncoupled legs of equal potential.
We then apply an optical potential offset $\Delta$ to one of the legs using light shaped by a digital micromirror device (DMD). 
We subsequently connect both legs by slowly lowering the lattice potential in the rung direction (for details see SI).
The offset between the two legs prevents the atoms from tunnelling along the rungs, such that we end up with roughly equally populated legs. 
Occupation readout is achieved with single-site spin and charge resolution~\cite{koepsell:2020}.
To highlight the influence of Pauli repulsion on hole pairing, we compare the mixD $\Delta \approx U/2$ case with the standard ladder system at $\Delta = 0$.
Both systems are realised in the strong rung-singlet regime, with $\Jp/J_\parallel = 21 (5)$ in the mixD case with enhanced spin exchange and $J_\perp/J_\parallel = 16 (3)$ in the standard ladders.
The mixD system is governed by the energy scales of rung spin exchange and leg tunnelling with $\tpl /\JpD = 0.7(1)$.
The strong rung coupling keeps the spatial extent of hole pairs small while staying in the regime where spin correlations and hole motion compete on a comparable energy scale.
For larger \tpl, the binding energy is even expected to grow~\cite{bohrdt:2021} but becomes harder to observe in systems of limited sizes due to increased pair sizes.

Fig.~\ref{fig:fig1}c shows the average density of the mixD system.
Doublon-hole fluctuations, dominantly from the energetically higher to the lower leg, as well as preparation imperfections, lead to a small residual density imbalance between the two legs (see SI).
For the data analysis (except Fig.~\ref{fig:fig1}c) we only take into account ladders without double occupancies, such that holes arising from doublon-hole fluctuations do not contribute to our observations.
Unless otherwise mentioned, we furthermore restrict the data to realisations with two to four holes per ladder and limit the occupation imbalance between the legs to one hole.

\begin{figure}[t]
\centering
\includegraphics{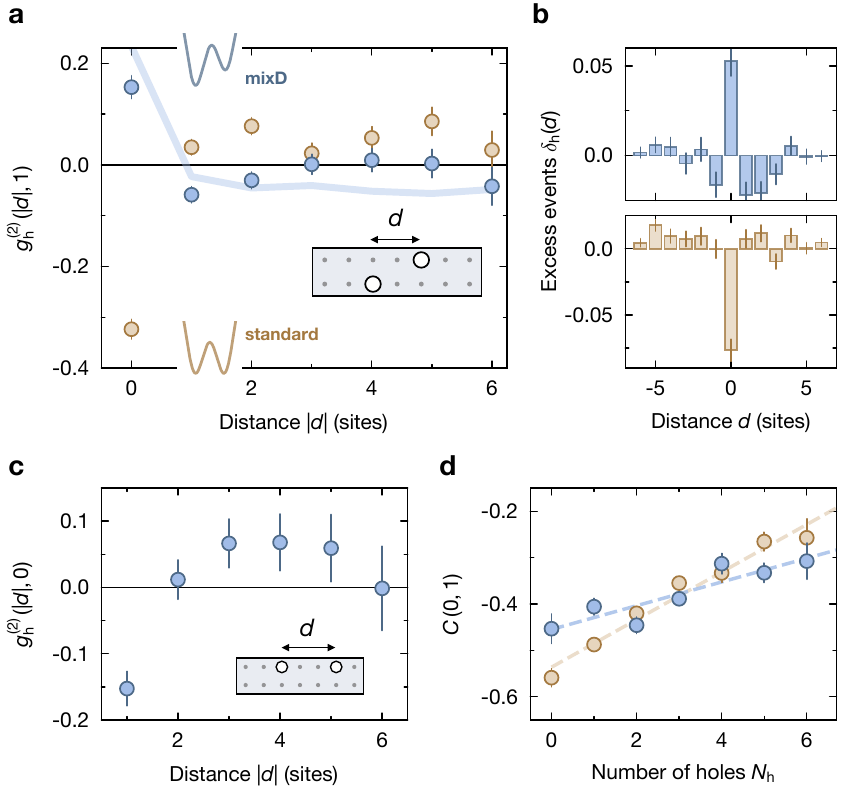}
\caption{
\textbf{Hole pairing in mixD versus standard ladders.}
\textbf{a}, Hole-hole correlator $\gh(d,1)$ between sites on opposite legs (as illustrated in the cartoon inset) for mixD (blue) and standard (brown) ladders with two to four holes per ladder.
The strong correlation at $d=0$ corresponds to two holes on the same rung. Correlations at this distance are strongly enhanced in the mixD system (pairing) and strongly suppressed in the standard ladders (repulsion).
The blue line is calculated using matrix product states (MPS) at finite temperature $\kB T=0.8\,\Jp$ and corrected by the experimental detection fidelity (see SI).
\textbf{b}, Excess events $\delta_{\mathrm{h}}(d)$ of the same data, i.e. likelihood to find holes at distance $d$ compared to the infinite temperature distribution.
\textbf{c}, Hole-hole correlation on the same leg $\gh(d,0)$ in the mixD system, showing that holes repel each other within the same leg.
A finite-size offset correction has been applied  to this subfigure (see SI).
\textbf{d}, Spin-spin correlation $C(0,1)$ for spins on the same rung depending on the number of holes in the system.
The lines represent linear fits.
The larger slope indicates that the spin order of the standard system (brown) is more strongly disturbed by holes than the spin order of the mixD system (blue), where paired holes leave the spin order largely unperturbed.
The errorbars denote one standard error of the mean (s.e.m) and are smaller than the marker when not visible. Errorbars in \textbf{a, c, d} are estimated using bootstrapping.
}
\label{fig:fig2}
\end{figure}

In order to probe the pairing of holes in our system, we evaluate the hole-hole correlator
\begin{equation}
\gh(\boldsymbol{r}) = \frac{1}{\mathcal{N}_{\boldsymbol{r}}}  \sum_{\boldsymbol{i}-\boldsymbol{j}=\boldsymbol{r}} \left(\frac{\langle \hat{n}^\mathrm{h}_{\boldsymbol{i}}\hat{n}^\mathrm{h}_{\boldsymbol{j}} \rangle}{\langle \hat{n}^\mathrm{h}_{\boldsymbol{i}}\rangle \langle \hat{n}^\mathrm{h}_{\boldsymbol{j}} \rangle} - 1 \right), 
\end{equation}
with normalisation $\mathcal{N}_{\boldsymbol{r}}$ the number of sites $\boldsymbol{i},\boldsymbol{j}$ at distance $\boldsymbol{r}$ and $\hat{n}^\mathrm{h}_{\boldsymbol{i}}$ the hole-density operator at position $\boldsymbol{i}$.
The function $\gh(\boldsymbol{r})$ is a connected two-point density correlator that is negative if the presence of a hole at position $\boldsymbol{i}$ makes it less likely to find a second hole at distance $\boldsymbol{r}$, and positive if it makes it more likely.
The correlator is bounded by $-1\le \gh(\boldsymbol{r}) \le (1/n_\text{h}-1)$ with hole density $n_\text{h}=N_\text{h}/2L$, where $N_\text{h}$ is the number of holes in the system.

By evaluating the correlation for holes in opposite legs $\boldsymbol{r}= (d,1)$, we observe a strong positive signal at distance $|d|=0$ in the mixD system of $\gh (0,1) = 0.15(2)$. This corresponds to two holes bunching on the same rung (see Fig.~\ref{fig:fig2}a).
The fast decrease of correlation for $|d|>0$ indicates that the holes are in a tightly bound state. 
We find a minimum at $|d|=1$, which we attribute to the effect of additional holes in the system. These holes are repelled from the hole pair (see also Fig.~\ref{fig:fig4}), leading to the weak modulation at larger distances which dominates over the extent of the hole pair at short distances.

In contrast, in the standard ladders of $\tp>\Jp$, a strong repulsion of holes from the same rung is the dominant feature leading to a negative $\gh(0,1)$ (see Fig.~\ref{fig:fig2}a). This shows that tightly bound pairs are energetically unfavourable and pairing, if at all, occurs at several sites distance.
At larger distances, we find a spatial modulation of the correlation consistent with a maximal mutual distance of holes along both the legs and rungs of the system.

The attraction (mixD) and repulsion (standard) of holes are also visible in the occurrences of hole distances.
In Fig.~\ref{fig:fig2}b we plot the histograms of holes found at a mutual distance $d$ as described by 
$\delta_{\mathrm{h}}(d) = \sum_{\boldsymbol{i}-\boldsymbol{j}=(d,1)} (\langle\hat{n}^\mathrm{h}_{\boldsymbol{i}}\hat{n}^\mathrm{h}_{\boldsymbol{j}}\rangle - n_\text{h}^2)$, where subtracting the global hole density $n_\text{h}$ removes the uncorrelated distribution.
These excess events $\delta_{\mathrm{h}}(d)$ can be interpreted as the likelihood for hole distance $d$ to occur beyond the probability of a random distribution.

The hole-hole correlator on the same leg (see Fig.~\ref{fig:fig2}c) shows the effect of hole mobility in the mixD system. 
It exhibits a minimum at nearest-neighbours caused by the Fermi-repulsion of holes due to the leg tunnelling $\propto t_\parallel$ and a broad maximum around $|d|=4$.
This is the largest mutual distance the two holes can assume without occupying the edge of the system, which is energetically expensive due to the hard walls blocking hole movement.

Since the pairing mechanism we are investigating is of magnetic origin, we characterise the interplay between holes and spins using the spin correlator
\begin{equation}
C(\boldsymbol{r}) = \frac{1}{\mathcal{N}_{\boldsymbol{r}}}  \sum_{\boldsymbol{\boldsymbol{i}}-\boldsymbol{\boldsymbol{j}}=\boldsymbol{r}}4\left( \langle \hat{S}^z_{\boldsymbol{i}}\hat{S}^z_{\boldsymbol{j}} \rangle_{\text{s}} -\langle \hat{S}^z_{\boldsymbol{i}}\rangle_{\text{s}}\langle\hat{S}^z_{\boldsymbol{j}} \rangle_{\text{s}} \right)
\label{eq:spincor}
\end{equation}
where $\langle\,\rangle_{\text{s}}$ denotes the expectation value for singly occupied sites.
In the doped mixD system we find strong nearest-neighbour spin correlations along the rung of $C(0,1) = - 0.38 (1) $, indicating a high singlet fraction, as well as a significant coupling of these bonds along the leg with nearest-neighbour spin correlations of $C(1,0) = - 0.10 (1)$.
Resolving the rung spin correlations through the number of holes in the system shows a decrease in correlation strength with growing hole number (see Fig.~\ref{fig:fig2}d).
This can be explained by unpaired holes breaking singlet bonds due to their mobility along the ladder.
The standard system, which does not display pairing, shows a more rapid loss of correlation strength compared to the mixD system, in which a significant fraction of holes is bound in pairs. This behaviour is directly related to the magnetic origin of pairing, as illustrated in Fig.~\ref{fig:fig1}.
The slight alternating behaviour of $C(0,1)$ at low hole numbers in the mixD system, in particular the strong spin correlations for two holes, is reminiscent of the low-temperature behaviour where rung pairs are dominant over thermal excitations and only odd hole numbers lead to unpaired holes (see SI).

\begin{figure}[t]
\centering
\includegraphics{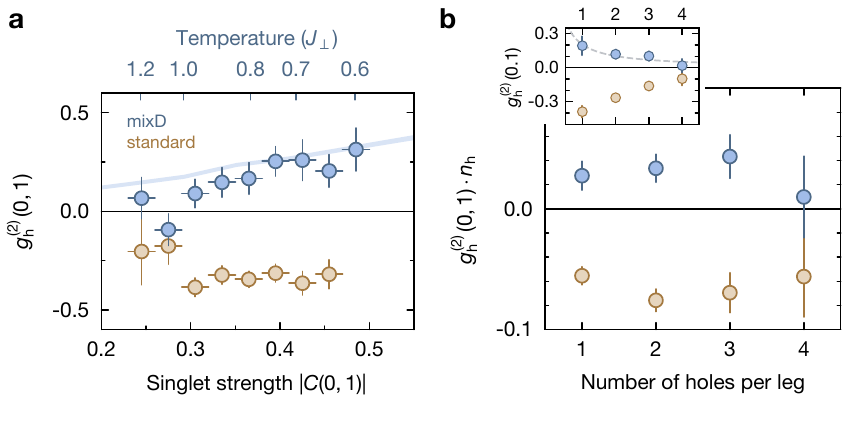}
\caption{\textbf{Temperature and doping dependence of hole pairing.} 
\textbf{a}, Rung hole-hole correlation $\gh(0,1)$ for the mixD (blue) and standard (brown) ladders binned by the rung spin correlations $C(0,1)$ of the system.
The temperature of the mixD system (top axis) is estimated by comparing the spin correlations (lower axis) to theoretical values.
The solid line is calculated using MPS and corrected by the experimental detection fidelity.
We see unbinding of pairs at low singlet strength, i.e. high temperature.
\textbf{b}, The hole correlator scaled with the hole density $\gh(0,1)\cdot n_\mathrm{h}$ depending on the number of holes per leg for the mixD (blue) and standard (brown) ladders. Within our errorbars, we find the hole binding to be independent of doping.
The inset shows the correlator $\gh(0,1)$, where the dotted line is a fit with the inherent $1/n_\text{h}$ scaling of the correlator.
Errorbars denote the bin width of the spin correlations (\textbf{a}) and the s.e.m. of the correlator (\textbf{a} and \textbf{b}).
}
\label{fig:fig3}
\end{figure}

We estimate the binding energy of the paired state in the mixD system from the experimental data, by comparing it to an analytically tractable effective Hamiltonian.
The approach is based on the assumption that the system is reasonably close to the uncoupled rung limit and the bound state can thus be described by two holes on the same rung (for details on the calculation, see SI).
Using the measured and detection fidelity corrected probability to find a rung pair and our estimated temperature of $\kB T=0.77(2)\,\JpD$ (see SI), we find an experimental binding energy of
\begin{equation*}
E_\text{b} =  0.82 (6)\, \JpD . 
\end{equation*}
This is consistent with DMRG calculations giving a binding energy of $E^{\mathrm{theo}}_\text{b} = 0.81\, \JpD$ and boosts the binding energy by an order of magnitude compared to standard ladders at the same interaction strength and with symmetric coupling (see also SI).

In order to gain a better understanding of the system dependencies, we probe the influence of temperature and doping on the rung pairing strength.
 In our experimental regime, nearest-neighbour spin correlations along the rung stand in strong direct relation to the temperature of the system and can therefore be used as an effective thermometer.
A mapping between the two is obtained by comparing average nearest-neighbour spin correlations to finite temperature MPS calculations (see SI).
We observe that the rung hole-hole correlation $\gh(0,1)$ increases with increasing spin correlation strength (see Fig.~\ref{fig:fig3}a), where the onset of pairing occurs in the experimental system around temperatures on the order of the spin exchange $\Jp$. 
The repulsion of the standard system remains mostly constant in our temperature regime, as it is governed by the energy scale $\tp \gg \Jp,\kB T$.

In order to elucidate the effect of doping on the pair binding, we adjust the correlator $\gh$ to compensate for the intrinsic density-dependent $1/n_\mathrm{h}$ scaling. 
The resulting correlator $n_\mathrm{h} \cdot \gh $ does not significantly change upon increasing the number of holes per leg (see Fig.~\ref{fig:fig3}b).
This is in agreement with a system of independent pairs for the mixD system.
The binding energy in the regime of tightly bound rungs mostly depends on the energy penalty for breaking a singlet, which is independent of density.
The rise in kinetic energy at higher doping due to the hard-core nature of the pairs is of minor influence in our parameter regime. 
For the standard system, hole repulsion is determined by the tunnelling strength \tp and is therefore also not significantly influenced by density.

\begin{figure}[t]
\centering
\includegraphics{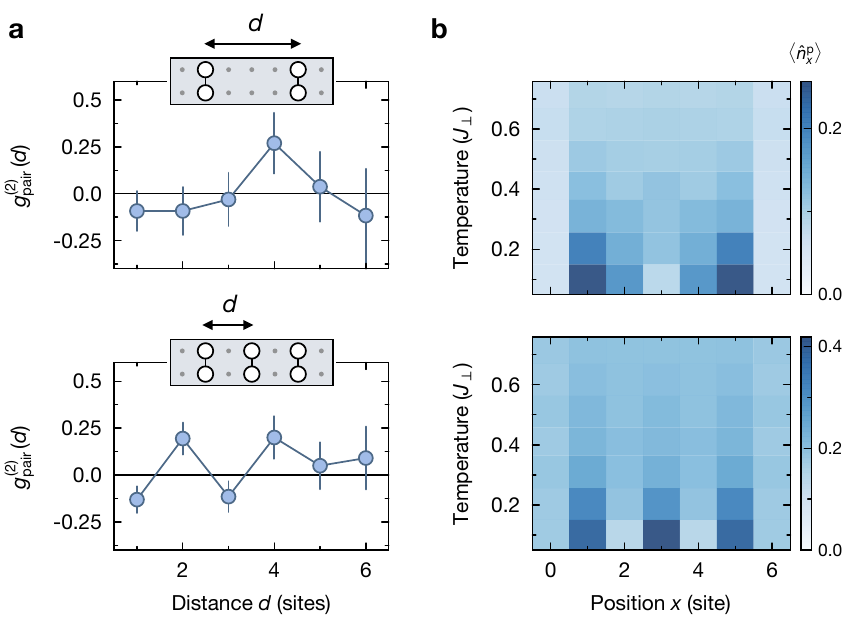}
\caption{\textbf{Distribution of rung hole pairs in the mixD system.} 
\textbf{a}, Measured pair-pair correlation $\gp(d)$ of rung hole pairs in the experimental system.
The upper plot shows the pair-pair correlation for 4-5 holes, i.e. up to two pairs, in the system.
The lower plot shows the pair-pair correlation for 6-7 holes, i.e. up to three pairs in the system.
A finite-size offset correction has been applied to the curves (see SI).
Errorbars were estimated using bootstrapping.
\textbf{b}, Theoretical (MPS) results for the density of rung pairs in the system for temperatures from $0.1\, \JpD$ to $0.7\,\JpD$.
The upper plot shows the pair density for four holes.
The lower plot shows the pair density for six holes.
In both cases, the pairs maximise their respective distance, while also avoiding the edge of the system.
}
\label{fig:fig4}
\end{figure}

We further investigate the behaviour of several pairs, as their interplay is a key aspect for the competition between superconductivity and charge (density) order~\cite{Fradkin2015,Qin2020}.
For simplicity, we identify the bound state as two holes occupying the same rung. 
We thus define the pair operator $\hat{n}^\mathrm{p}_x$ which is equal to $1$ if both sites of rung $x$ are occupied by a hole, and $0$ otherwise.
Pair interactions are then quantified by the pair-pair correlator
\begin{equation}
\gp(d) = \frac{1}{\mathcal{N}_{d}}  \sum_{x} \left(\frac{\langle \hat{n}^\mathrm{p}_{x\vphantom{d}}\hat{n}^\mathrm{p}_{x+d} \rangle}{\langle \hat{n}^\mathrm{p}_{x\vphantom{d}}\rangle \langle \hat{n}^\mathrm{p}_{x+d} \rangle} - 1 \right),
\label{eq:pair_correlator}
\end{equation}
in analogy with the hole-hole correlator $\gh$. 
We evaluate $\gp$ on a sub-set of our data in which at least two pairs are present in the system.
In the case of ladders containing 4 - 5 holes (Fig.~\ref{fig:fig4}a), we find a peak in the correlator at distance $|d| = 4$, while it peaks at distances $|d| = 2$ and $|d| = 4$ for the ladders containing 6 - 7 holes.
The pairs thus arrange in a spatial structure for which they maximise both their mutual distance and the distance to the edges of the system.
Two holes together can delocalise in the system, but two pairs display hard core interaction leading to mutual repulsion.

To investigate these pair interactions numerically,
we perform finite temperature MPS simulations on a system with the same parameters as in the experiment ($L = 7$, $\tpl/\JpD = 0.7$).
Here, pair interaction is directly revealed by the density of pairs $\langle \hat{n}^\mathrm{p}_x \rangle$ which shows strong spatial dependence in the low temperature regime (see Fig.~\ref{fig:fig4}b). This visibility in the pair density is a direct consequence of our open boundary conditions, as the presence of sharp edges fixes the phase of the density modulation.
In systems with four holes, bound pairs primarily sit next to the edge sites, corresponding to the peak at $|d| = 4$ that we observe in the pair correlator evaluated on the experimental data.
Similar behaviour is observed on systems with six holes, where pairs keep a distance of $d=2$ sites to each other.
Such a modulation of the pair distribution is reminiscent of Friedel oscillations of indistinguishable fermions near an impurity~\cite{friedel:1952}, as well as of charge-density-waves~\cite{ white:2002}. Larger systems are needed to distinguish between these signatures.

In this work, we demonstrated the direct observation of hole pairing in a quantum gas microscope setting.
We have realised a paradigmatic model that reaches hole binding at high temperatures close to the spin-exchange energy and small pair size by engineering doped, mixed-dimensional fermionic ladders.
We confirm that the suppression of Pauli repulsion enables the formation of a bound state, allowing us to experimentally investigate its mechanism based on hole motion and magnetic correlations. 
Further analysis revealed the thermal unbinding of pairs at lower spin correlation strength.
Finally, we have seen signs for significant mobility of the bound pairs through their repulsive interaction which resembles hard core particles in a 1D system.
Our experiment leads the way to explore even higher binding energies in larger systems at $\tpl>\Jp$ and stripe formation at higher leg number~\cite{grusdt:2018a}. Our mixed-dimensional setting can furthermore be readily extended to higher dimensions using bilayer quantum gas microscopes~\cite{preiss:2015,hartke:2020, koepsell:2020}, where highly mobile pairs are expected to form a quasi-condensate at high critical temperature~\cite{bohrdt:2021}.
Our results thus pave the way to the measurement of collective phases of bound pairs like crystallisation and superfluidity, and shed light on their competition~\cite{Fradkin2015,Qin2020,Rey2009,Trebst2006}.

\textbf{Acknowledgements:} We thank T. Giamarchi for fruitful discussions. We thank A. Kantian for fruitful discussions and critical reading of the manuscript. This work was supported by the Max Planck Society (MPG), the European Union (FET-Flag 817482, PASQUANS), the Max Planck Harvard Research Center for Quantum Optics (MPHQ), the German Federal Ministry of Education and Research (BMBF grant agreement 13N15890, FermiQP) and under Germany's Excellence Strategy -- EXC-2111-390814868.
T.C. acknowledges funding from the Alexander v. Humboldt Foundation.
F. G. acknowledges funding from the European Research Council (ERC) under the European Union’s Horizon 2020 research and innovation programme (Grant Agreement no 948141) — ERC Starting Grant SimUcQuam. A. B acknowledges funding from the NSF through a grant for the Institute for Theoretical Atomic, Molecular, and Optical Physics at Harvard University and the Smithsonian Astrophysical Observatory. E.D. acknowledges funding from the ARO (grant number W911NF-20-1-0163) and AFOSR-MURI: Photonic Quantum Matter award FA95501610323.


\textbf{Competing interests}: The authors declare no competing interests.
\bigskip

\bibliographystyle{naturemag}

\bigskip

\clearpage
\newpage
\makeatletter
\renewcommand{\thefigure}{S\@arabic\c@figure}
\makeatother
\setcounter{figure}{0}
\setcounter{table}{0}

\section*{Supplementary Information}
\subsection*{Experimental sequence}
In each experimental run, we prepare a cold atomic cloud of $^6$Li in a balanced mixture of the lowest two hyperfine states $\ket{F=1/2, m_F = \pm 1/2}$.
For evaporation, we confine the cloud in a single layer of a staggered optical superlattice along the $z$-direction with spacings $a_\mathrm{s} = \SI{3}{\micro\meter}$ and $a_\mathrm{l} = \SI{6}{\micro\meter}$, and depths $V_\mathrm{s} = 51\, E_\mathrm{R}^\mathrm{s}$ and $V_\mathrm{l} = 120\,E_\mathrm{R}^\mathrm{l}$, where $E_\mathrm{R}^{\alpha} = h^2/(8Ma_{\alpha}^2)$ denotes the recoil energy of the respective lattices ($\alpha = \mathrm{s}, \mathrm{l}$), and $M$ is the mass of an atom.
The atoms are harmonically confined in the $xy-$plane and the evaporation is performed by ramping up a magnetic gradient along the $y$-direction (see~\cite{koepsell:2020}).
The final atom number is adjusted via the evaporation parameters.

\begin{figure}[ht!]
\begin{center}
\includegraphics{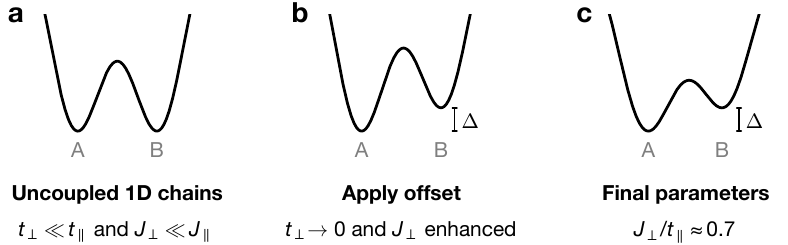}
\caption{
\textbf{Preparation sequence for mixD systems.}
\textbf{a}. We first prepare nearly uncoupled 1D chains in which the leg tunnelling exceeds the rung coupling.
\textbf{b}. While the legs are decoupled, we apply the offset $\Delta$ to one leg of the ladder.
\textbf{c}. The final parameters are reached by ramping down the leg coupling and ramping up the rung coupling.
There, the potential offset $\Delta$ between legs prevents tunnelling from one leg to the other.
Note that in the final configuration $\Jp \gg \Jpl$.
}
\label{fig:S1}
\end{center}
\end{figure}

We adiabatically load the cloud into an optical lattice in the $xy-$plane with spacings $a_{\mathrm{x}} = \SI{1.18}{\micro\meter}$ and $a_{\mathrm{y}} = \SI{1.15}{\micro\meter}$.
Simultaneously, we apply a repulsive potential using a digital micromirror device (DMD), which both compensates for the harmonic confinement of the Gaussian shaped lattice beams, and shapes the system into a geometry of four $2\times7$ ladders following the procedure described in~\cite{sompet:2021}. The DMD is furthermore used to apply a spin independent optical potential offset $\Delta$ between sites along the rung direction.
The loading is performed in three steps (see Fig.~\ref{fig:S1}).
(a) A first stage, in which the two legs of each ladder are nearly disconnected, is reached by ramping the lattice depths to $V_{\mathrm{x}} = 20\,E_{\mathrm{R}}^{\mathrm{x}}$ and $V_{\mathrm{y}} = 3\,E_{\mathrm{R}}^{\mathrm{y}}$ in \SI{100}{\milli\s}.
(b) The optical potential offset $\Delta$ (see \emph{``Potential offset calibration''}) is applied to one leg of each ladder by instantaneously ($< \SI{20}{\micro\second}$) switching the pattern of the DMD.
(c) The lattice depths are ramped linearly to their final values $V_{\mathrm{x}} = 6\,E_{\mathrm{R}}^{\mathrm{x}}$ and $V_{\mathrm{y}} = 12\,E_{\mathrm{R}}^{\mathrm{y}}$ in \SI{100}{\milli\s}.

Interactions between the atoms are set by the $s$-wave scattering length $a_{s}$, which we adjust by applying a magnetic field close to the broad Feshbach resonance of $^{6}$Li around \SI{830}{\Gauss}.
The scattering length is increased from $a_s \approx 230\,a_\text{B}$ during evaporation, with $a_\text{B}$ being the Bohr radius, to $a_s \approx 1310a_{\mathrm{B}}$ in the final configuration. 
The resulting system is described by the Fermi-Hubbard model and an additional potential offset $\Delta$. 
Our parameters are the repulsive on-site interaction $U = h\times\SI{4.29(10)}{\kHz}$, tunnelling $\tplbare = h\times\SI{78(10)}{\Hz}$ and $\tpbare = h\times\SI{303(23)}{\Hz}$ and the offset $\Delta \approx 0.5\,U$ or $\Delta = 0 $ depending on the configuration (mixD or standard).
Since $U/\tpbare,U/\tplbare \geq 14$, the system can be effectively described by the $t-J$ model (see also \emph{``From the Fermi-Hubbard to the $t-J$ model''}).
Note that we use $\tpbare,~\tplbare$ for the tunnelling parameters in the Hubbard model, and $\tp,~\tpl$ in the $t-J$ model.
Along the legs, the tunnel coupling is independent of $\Delta$ and is $\tpl = \tplbare =  h\times\SI{78(10)}{\Hz}$, yielding a spin exchange of $\Jpl = h\times5.7(1.5)\,\si{\Hz}$.
Along the rungs, the mixD system ($\Delta/U \approx 0.5$) yields $\tp = 0$ and an enhanced spin exchange $\Jp = h\times\SI{114(42)}{\Hz}$.
Without the potential offset, i.e. in the standard system ($\Delta = 0$), tunnelling is unaffected, leading to $\tp= \tpbare = h\times\SI{303(23)}{\Hz}$ and $\Jp = h\times\SI{86(13)}{\Hz}$.

\subsection*{Potential offset calibration}

\begin{figure}[t]
\begin{center}
\includegraphics{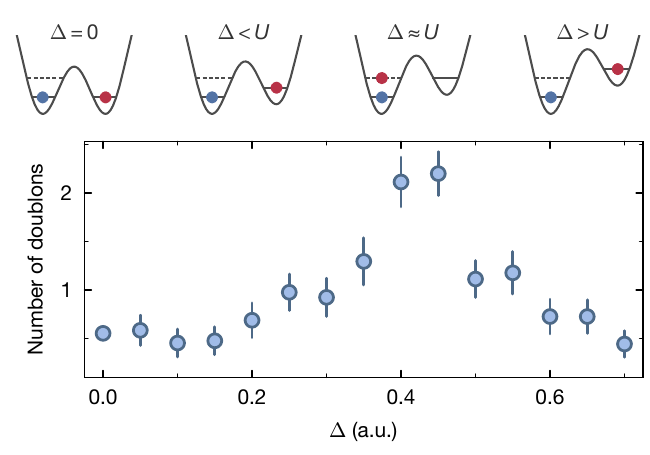}
\caption{
	\textbf{Calibration of the optical potential offset.}
	The experimental sequence is run for different value of $\Delta$ in a regime close to unit occupancy of the lattice.
	tunnelling from one leg to the other is suppressed as long as $|\Delta - U| > 0$.
	When $\Delta \sim U$, tunnelling is possible, and an increased number of doublons in the system is measured.}
\label{fig:S2}
\end{center}
\end{figure}

We realise the mixed-dimensional system by applying a local spin independent light shift $\Delta$ to one of the legs on each ladder. The amplitude is directly proportional to the light intensity, which is controlled by the DMD.
Calibration of the offset is performed by running the experimental sequence described above for different light intensities, and measuring the density of doubly-occupied sites (\emph{doublons}) in the system.
An increase of doublons is seen when $\Delta \approx U$, i.e. when the lowest band of the upper leg becomes resonant with the interaction band of the lowest leg (Fig.~\ref{fig:S2}).
This calibration was repeated several times throughout data collection, with typical shift of the doublon peak of $\sim 10\,\%$.
We attribute these calibration differences to drifts in the beam shape of the light that is sent to and diffracted from the DMD, yielding an overall estimation of the uncertainty on $\Delta$ of about $\pm \SI{15}{\percent}$. Such uncertainty in $\Delta$ is not critical for realising a mixed-dimensional setting and mostly influences the value of \Jp. 

\subsection*{Suppression of rung tunnelling}
\begin{figure}[t]
\begin{center}
\includegraphics{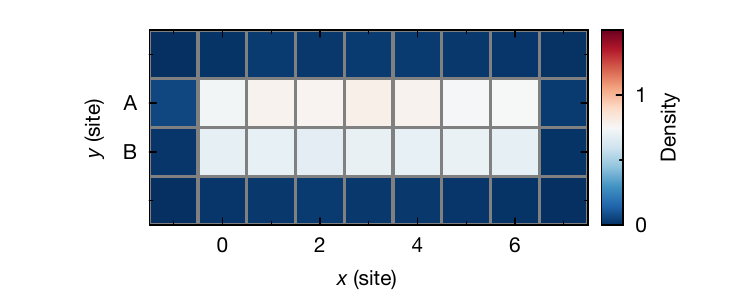}
\caption{
\textbf{Density of the mixD system without doublons.}
The density of the mixD system, where only ladders without double occupancies are taken into account.}
\label{fig:S12}
\end{center}
\end{figure}

The potential offset $\Delta \gg \tpbare$ between the two legs shifts the energy levels between neighbouring sites and thus suppresses tunnelling along the rungs~\cite{grusdt:2018a}. 
Doublon-hole pairs, however, become biased in the mixD system and appear predominantly as double occupancy on the leg with lower potential and corresponding empty site on the upper leg. While in the standard system doublon-hole pairs along the rung appear with probability $\sim (\tpbare/U)^2$~\cite{hartke:2020}, the potential offset in the mixD case lowers the energy difference between the singly occupied state and the doublon-hole pair to $U-\Delta$. This effect can be seen in the density of the mixD system in Fig.~\ref{fig:fig1}c. In Fig.~\ref{fig:S12}, we plot the same data after removing ladders containing double occupancies. The density imbalance mostly disappears, indicating minimal tunnelling during preparation.

\subsection*{Detection}
The data presented in the main text originates from two types of measurements: (a) charge-resolved and (b) spin-charge-resolved measurements.
In both cases, the detection procedure starts by ramping the $xy-$lattices to $43\,E_\mathrm{R}^{\mathrm{xy}}$ within $\SI{250}{\micro\second}$, effectively freezing the occupation configuration.
In the case of spin-resolved measurements (b), a Stern-Gerlach sequence separates the two spin species into two neighbouring planes of the vertical superlattice, which are then separated by $\SI{21}{\micro\meter}$ from one another using the charge pumping technique described in~\cite{koepsell:2020}.
Finally, fluorescence images are taken using Raman sideband cooling in our dedicated pinning lattice with an imaging time of $\SI{1}{\second}$~\cite{omran:2015}.
For a charge-only measurement (a), only one plane is populated by atoms, while in the case of a fully-resolved measurement (b) two planes are populated by the two different spin species.
The fluorescence light of the atoms is then collected through a high-resolution objective and imaged onto an EMCCD camera.
For a fully spin-resolved measurement (b), the fluorescence of both planes is collected simultaneously and imaged on the camera, allowing to reconstruct the atomic distribution of both spins with a single exposure.
A charge-only measurement only allows to reconstruct the atomic configuration, without any spin information.

The imaging technique and the pumping procedure both impact our overall detection fidelity.
The imaging fidelity, which takes into account atom losses and atom displacement during the imaging procedure, is estimated by comparing two consecutive fluorescence pictures of the same atomic distribution, and we obtain an average imaging fidelity $\mathcal{F}_{\mathrm{I,a}} = \SI{98.7(1)}{\percent}$ ($\mathcal{F}_{\mathrm{I,b}} = \SI{98.2(2)}{\percent}$) per atom for charge-only (resp. full-spin-charge) resolution.
The pumping fidelity is estimated by comparing the average number of atoms detected after pumping to the average number of atoms before pumping, and we obtain an average pumping fidelity of $\mathcal{F}_{\mathrm{P}} = \SI{97.6(1)}{\percent}$, taking into account the slight discrepancy between $\mathcal{F}_{\mathrm{I,a}}$ and $\mathcal{F}_{\mathrm{I,b}}$.
We deduce an overall detection fidelity of $\mathcal{F}_\mathrm{a} = \mathcal{F}_{\mathrm{I,a}} = \SI{98.7(1)}{\percent}$ ($\mathcal{F}_\mathrm{b} = \mathcal{F}_{\mathrm{I,b}}\mathcal{F}_{\mathrm{P}} = \SI{95.8(1)}{\percent}$) in the case of charge-only (full-spin-charge) resolution.

\subsection*{Data statistics}

We have taken $\sim$19\,000 experimental shots, iterating between mixD $\Delta \approx U/2$ and standard $\Delta = 0$.
61\,\% of the shots have charge-only resolution, 39\,\% have full spin and charge resolution.

\begin{figure}[t]
\begin{center}
\includegraphics{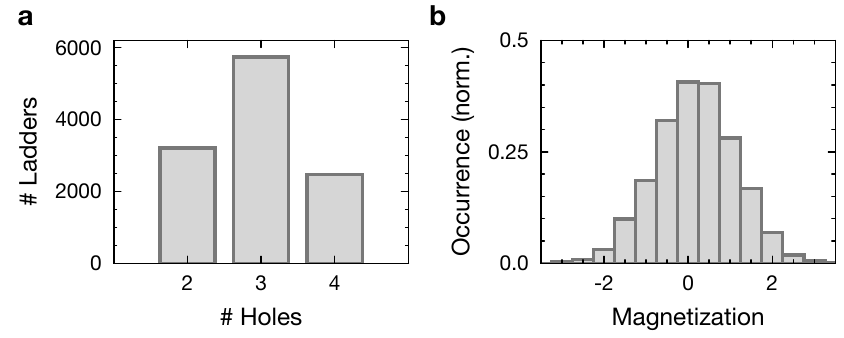}
\caption{
	\textbf{Hole and magnetisation statistics.}
	Experimental distribution of holes per ladder (\textbf{a}) and total magnetisation (\textbf{b}) for the data shown in Fig.~\ref{fig:fig2}a-c, and Fig.~\ref{fig:fig3}a.
}
\label{fig:S3}
\end{center}
\end{figure}

The ladders are very sensitive to small drifts in the DMD pattern relative to the lattice sites.
We thus keep track of the ladder potential by continuous automatic evaluation of the charge distribution and automatic feedback to the DMD pattern.
If the average leg-to-leg occupation imbalance of untilted ladders exceeds 2 holes, we dismiss the respective set of data due to the uncontrolled drift in the potential.
For data analysis we then only take into account ladders without double occupancies and with a leg-to-leg occupation imbalance of maximally one hole. This leaves us with more than 24\,000 individual ladders, about half of which contain between two and four holes (see Fig.~\ref{fig:S3}a). The majority of ladders shows a magnetisation $|M^z| < 2$, with $M^z=\sum_{\boldsymbol{i}}\hat{S}^z_{\boldsymbol{i}}$ (see Fig.~\ref{fig:S3}b).
Figures and values given in the main text, unless otherwise mentioned, are filtered for two to four holes.

\subsection*{Numerical simulations - DMRG}

We numerically simulate the $t-J$ model, Eq.~\eqref{eqn:tJ} in the main text, using matrix product states (MPS). For the mixed-dimensional ($\tp=0$) case, we set the parameters to $J_{\parallel}/J_\perp = 0.047$, $t_{\parallel}/J_\perp = 0.7$. In the standard ($\tp>0$) case, the parameters are $J_{\parallel}/J_\perp = 0.06$, $t_{\parallel}/J_\perp = 0.9$, and $t_\perp/J_\perp = 3.57$. This corresponds to the $t-J$ model derived from a Fermi-Hubbard model with $U/t_\perp = 14.16$, $t_{\parallel}/t_\perp = 0.26$, and, in the mixed-dimensional case, $\Delta/U = 0.5$.
We use the TeNPy package~\cite{Hauschild2018,Hauschild2019} to perform the MPS simulations. In order to simulate systems at finite temperature, we use the purification method~\cite{Verstraete2004,Schollwoeck2011}, where the Hilbert space is enlarged by an auxiliary site $a(i)$ per physical site $i$. The finite temperature state of the physical system is obtained by tracing out the auxiliary degrees of freedom. We start from an infinite temperature state, in which the physical and auxiliary degrees of freedom on each site are maximally entangled. In particular, we implement an entangler Hamiltonian~\cite{Nocera2016} to prepare the infinite temperature state of the $t-J$ model. We work in the grand canonical ensemble and thus introduce a chemical potential $\mu$ to control the average number of holes in the system. Starting from the infinite temperature state, we then use the $W^{II}$-time-evolution method~\cite{Zaletel2015} to perform imaginary-time evolution up to the desired temperature. 

In order to directly compare to the experimental data, we directly sample snapshots from the matrix product state using the perfect sampling algorithm~\cite{Ferris2012}. 
In the evaluation of the snapshots, we account for the experimental detection fidelity by randomly placing artificial holes in the MPS snapshots according to our detection fidelity. We then apply the same filters regarding hole number and occupation imbalance as for the experimental data and model the hole number distribution of the experimental data (see Fig.~\ref{fig:S3}a) by weighting the snapshots accordingly.

For ground state simulations, for example to obtain the binding energies, we use the DMRG algorithm and work in a fixed $S_z^{\mathrm{tot}}$ as well as particle number sector. 

\subsection*{From the Fermi-Hubbard to the $t-J$ model}

The Fermi-Hubbard model
\begin{equation*}
    \mathcal{H} = -\sum_{\langle i , j\rangle,\sigma}  - \tilde{t}_{ij}\,\left( \hat{c}^{\dagger}_{i,\sigma}\hat{c}^{\vphantom{\dagger}}_{j,\sigma} + \text{h.c.}\right) \,+ U \sum_{i}\hat{n}_{i,\uparrow}\hat{n}_{i,\downarrow}
\end{equation*}
contains a hopping term and (repulsive) on-site interaction. In the limit of large interactions $U \gg t$, where $U/\tilde{t}$ needs to be large enough to be well into the Mott-insulating regime, double occupancies are suppressed. An expansion to leading order in $\tilde{t}/U$ yields several terms, including the $t-J$ model (\ref{eqn:tJ}) with $J=4\tilde{t}^2/U$ and thus $t\gg J$. The expansion furthermore yields terms on the order of $t^2/U$~\cite{auerbach:1994} describing next-nearest-neighbour hopping via a (virtual) double occupancy, in analogy to the spin exchange term.

For our mixD system the only term arising is $\sim \tpl^2/U \ll \Jp,\tpl$, which is much smaller than the relevant energy scales in the system and can thus be omitted.
For the standard system there are more possible combinations of processes, like $\sim \tpl\tp/U $, which is much larger than the process including only $\tpl$, but the system is still dominated by $\tp,\ \tpl,\ \Jp$.
In the parameter regime where $\tpl \gg \Jp$, however, this term becomes increasingly important such that the Fermi-Hubbard system can eventually not be approximated by the $t-J$ model of Eq. (\ref{eqn:tJ}).
This explains discrepancies found in the literature between binding energies calculated in $t-J$ ladders~\cite{white:1997a,Chernyshev1998} and in Fermi-Hubbard ladders~\cite{Karakonstantakis2011} in the same parameter regimes.

\subsection*{Temperature estimation}
\begin{figure}[t]
\begin{center}
\includegraphics{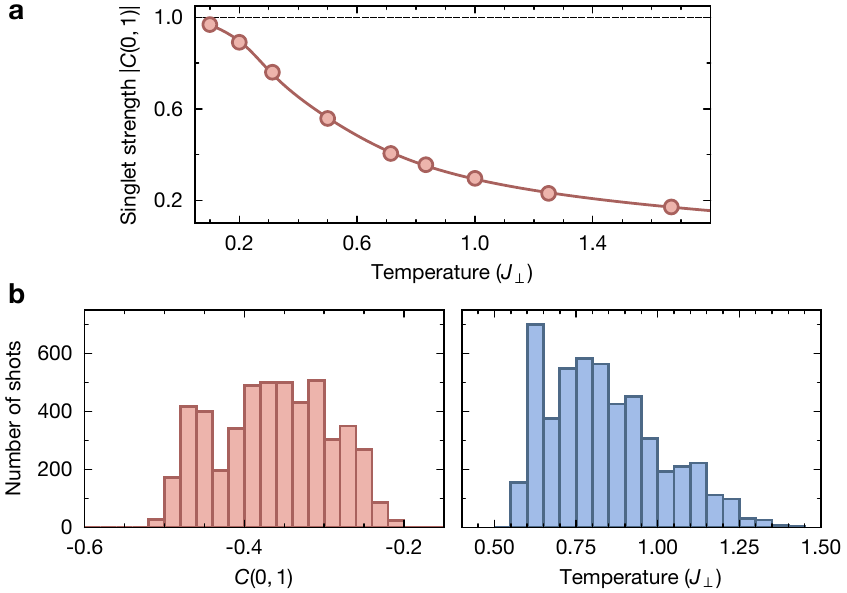}
\caption{
    \textbf{Temperature estimation.}
	\textbf{a,} Singlet strength vs. temperature.
	The calibration of temperature is performed using MPS data containing two to four holes.
	\textbf{b,} Experimental singlet strength and inferred temperature distributions.
	We evaluate our rung spin correlations $C(0,1)$ on the mixD system, using a time window of about $\SI{24}{\hour}$.
	The temperature is extracted from $C(0,1)$ using the MPS simulation (panel \textbf{a}).
}
\label{fig:S7}
\end{center}
\end{figure}

We estimate the temperature of our system by comparing the measured rung spin correlations $C(0,1)$ as defined in Eq.~(\ref{eq:spincor}) to the values calculated from MPS snapshots (Fig.~\ref{fig:S7}a).
We find that our average rung spin correlations of $C(0,1) = -0.38(1)$ for two to four holes correspond to a temperature of $\kB T=0.77(2)\Jp$. 

Our data is however not well described by a single spin correlation value, as we see variations both in time and across the four simultaneously realised ladders. 
The temperature estimation for the full dataset is therefore an average, while the data can contain features of both lower and higher temperatures. 
One reason for temperature variations are drifts in the apparatus on a time scale of days, affecting in particular the evaporation stage, which sets the global temperature.
Another reason is the potential shaping, which distributes entropy between the four ladders and the surrounding bath.
We thus attribute a temperature to each ladder (out of the four ladders we realise simultaneously) and each point in time by averaging the spin correlations of a time window of about $\pm \SI{12}{\hour}$.
The resulting spin correlation and temperature distribution are shown in Fig.~\ref{fig:S7}b.

\subsection*{Correlation functions}
Evaluating correlators in finite sized systems with fixed particle number leads to finite-size offsets, due to self correlation of the particles. For our purpose we have to distinguish two cases. 
For correlations between different legs, e.g. the rung hole correlation $\gh(0,1)$, self correlation does not cause problems. In the mixD case holes can not move from one leg to the other, such that finding a hole in leg $A$ does not influence the number of holes in leg $B$. In the standard case holes are mobile between the legs, but the focus of the analysis still lies on holes in opposite legs, since we select the data for low occupation imbalance. The correlations are thus not influenced by self correlation.
Correlations within the same leg are, however, strongly affected by finite-size offsets. 
We correct for these offsets using
\begin{equation*}
\gho(d,0) = \frac{1}{\mathcal{N}_{d}}  \sum_{\boldsymbol{i}-\boldsymbol{j}=(d,0)} \left(\frac{\langle \hat{n}^{\mathrm{h}}_{\boldsymbol{i}}\hat{n}^{\mathrm{h}}_{\boldsymbol{j}} \rangle}{\langle \hat{n}^{\mathrm{h}}_{\boldsymbol{i}}\rangle \langle \hat{n}^{\mathrm{h}}_{\boldsymbol{j}} \rangle}\frac{N_\text{l}}{N_\text{l}-1} - \frac{L}{L-1}\right),
\end{equation*}
where $N_\text{l}$ is the number of holes in the leg and $L$ is the length of the leg. The same offset correction is applied to the pair correlator of Eq. \eqref{eq:pair_correlator} in the main text.
The offset correction applies a distance independent correction and thus affects the overall value, but not the shape of the curve.

\subsection*{Binding energy}

We estimate the binding energy of holes from the measured correlation $g^{(2)}_h(0,1)$ of two holes on the same rung.
To this end, we simplify the mixed-dimensional $t-J$ Hamiltonian [Eq.~(\ref{eqn:tJ}) in the main text] by neglecting the two smallest energy scales $J_{\parallel}, t_\parallel$.
This is partly justified by the fact that both are below the estimated temperature $T$ of the experiment.

As a result, the Hamiltonian completely decouples into individual rungs and we can exactly diagonalise the latter.
Then, as detailed below, we perform a canonical calculation of the entire system, with exactly one hole on each of the two legs of length $L$.
From the known temperature $T$ and the rung super-exchange $J_\perp$ we obtain a direct relation between the binding energy $E_{\rm b}$ and the rung-correlation function $g^{(2)}_h(0,1)$:
\begin{equation}
    E_{\rm b} = - \beta^{-1} \log \left[ \frac{\left( 1 + 3 e^{-\beta J_\perp} \right) \left( 1 - \frac{g^{(2)}_h(0,1)}{L-1} \right) }{4 \left( 1 + g^{(2)}_h(0,1) \right)} \right] 
    \label{eqEbindEstimate}
\end{equation}
where $\beta = 1/(k_B T)$.

In order to use the measured correlation value given in Fig.~\ref{fig:fig2}a, we have to eliminate the density dependence of the \gh correlator. Using the insights of Fig.~\ref{fig:fig3}b, we scale the hole correlator with the hole density $n_\mathrm{h}$. 
Using the scaled correlator $n_\mathrm{h}\cdot\gh$ and the above formula with the experimentally estimated values for $k_B T / J_\perp = 0.77(2)$, we obtain the estimate for the binding energy $E_{\rm b} =0.82(6) \,J_\perp$ stated in the main text. The error derives from the error on the experimental value and the error on the temperature estimation. If we use the measured hole correlation for exactly two holes in the system (Fig.~\ref{fig:fig3}b), as it is used in the above derivation of (\ref{eqEbindEstimate}), we obtain a binding energy of $E_{\rm b} =0.79(9) \,J_\perp$. Both calculations yield results in very good agreement with the theoretical prediction from DMRG at $L=7$ of $E_{\rm b}^{\mathrm{theo}} =0.81 \,J_\perp$. In the limit of large systems the DMRG gives a binding energy of $E_{\rm b}^{\mathrm{\infty}} =0.78\, J_\perp$, demonstrating that our system with its tightly bound pairs provides a good approximation to the physics in larger systems.

\begin{figure}[t]
\begin{center}
\includegraphics{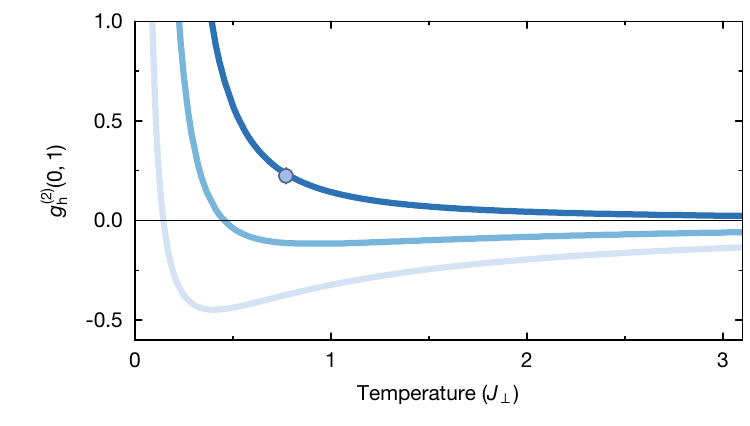}
\caption{
	\textbf{Hole correlation strength $\gh(0,1)$ as predicted by our simplified analytic model.}
	The rung correlation $\gh(0,1)$ as predicted by our model (Eq.~\eqref{eqEbindEstimate}) is plotted against temperature for a binding energy of $E_\mathrm{b} = 0.3 \Jp$ (light blue), $E_\mathrm{b} = 0.5 \Jp$ (blue) and $E_\mathrm{b} = 0.8 \Jp$ (dark blue). For our experimental system of $\kB T= 0.77(2) \Jp$, only the high binding energy yields a positive $\gh(0,1)$. The marker indicates the experimental result.
}
\label{fig:S11}
\end{center}
\end{figure}

In the remainder of this section, we explain the simplified model used here in more detail and derive from it Eq.~\eqref{eqEbindEstimate}. As mentioned in the beginning, we neglect the smallest energy scales $t_\parallel$ and $J_\parallel$. The eigenstates of each decoupled rung therefore become the two-hole state $\ket{{\rm hh}}$, the four spin-hole states $\ket{{\rm sh},y,\sigma}$ with leg index $y=0,1$ and spin index $\sigma=\uparrow, \downarrow$, the spin-singlet state $\ket{{\rm S}}$ and the three spin-triplet states $\ket{{\rm T},m}$ with $m=-1,0,1$. The corresponding eigenenergies are $\epsilon_{\rm hh}=V$, $\epsilon_{\rm sh}=\epsilon_{\rm T} = 0$ and $\epsilon_{\rm S}=-J_\perp$. 
Note that we allowed for a variable energy $V$ of the ${\rm hh}$ state. For $t_\parallel = J_\parallel =0 $ we know that $V=0$, however for small but non-zero couplings $t_\parallel, J_\parallel$ a non-zero renormalisation of $V \neq 0$ can be expected. The strength of $V$ can be calculated perturbatively~\cite{bohrdt:2021a}, but we treat it as a free parameter here which allows us to go beyond a perturbative analysis. 

We start by defining the binding energy of the simplified model in the thermodynamic limit $L \to \infty$. To this end we compare the ground state energy of a system with two independent holes, $2 ( E_{\rm 1h} - E_{\rm 0h})= 2 J_\perp$, with the ground state energy of a system with one pair of bound holes, $E_{\rm 2h} - E_{\rm 0h} = V + J_\perp$; both are measured relative to the undoped ground state, $E_{\rm 0h} = - L J_\perp$. The binding energy is then defined as
\begin{equation}
    E_{\rm b} = 2  E_{\rm 1h} - E_{\rm 0h} - E_{\rm 2h} = J_\perp - V.
\end{equation}
For $E_{\rm b} >0$ ($E_{\rm b}<0$) the two-hole ground state is paired (unpaired).

To derive Eq.~\eqref{eqEbindEstimate} we perform a canonical calculation with exactly one hole per leg. The probability for finding both holes on the same rung anywhere in the system becomes $p_{\rm hh} = L e^{-\beta E_{\rm hh}} Z_{\rm S}^{L-1}/Z$, where we defined the spin $Z_{\rm S} = e^{- \beta E_{\rm S}} + 3 e^{-\beta E_{\rm T}}$ and total partition functions $Z=L e^{- \beta E_{\rm hh}} Z_{\rm S}^{L-1} + 4 L (L-1) e^{-2 \beta E_{\rm sh}} Z_{\rm S}^{L-2}$. By the definition of the $g^{(2)}$ function provided in the main text, we obtain the relation
\begin{equation}
    g^{(2)}_h(0,1) = \frac{p_{\rm hh}/L}{(1/L)^2} - 1
\end{equation}
within our model. Simplifying this expression and solving for $E_{\rm b}$ finally yields Eq.~\eqref{eqEbindEstimate}.

This formula furthermore reveals the dependence of the rung hole correlator on temperature and binding energy (Fig.~\ref{fig:S11}). For lower binding energies of around $0.5 \Jp$, the $\gh(0,1)$ correlator turns negative at the temperature of the experimental system of $T\approx 0.8 \Jp$, despite the ground state being a paired state and the correlator displaying positive values at lower temperature. This is due to the higher entropy of the unbound state, having four possible spin configurations for each hole configuration, and the minimisation of free energy at finite temperature.

\begin{figure}[t]
\begin{center}
\includegraphics{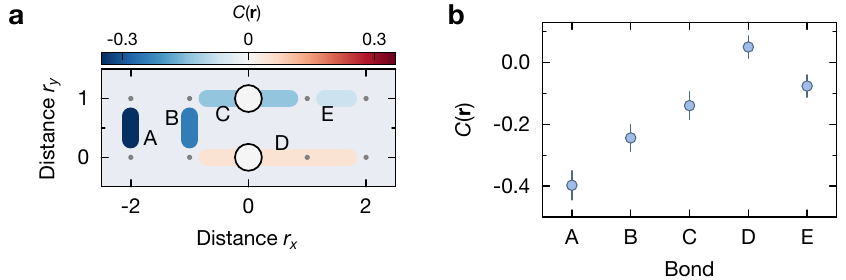}
\caption{
	\textbf{Spin correlations around a hole pair.}
	\textbf{a}, Map of spin correlations around a hole pair in the mixed-dimensional system. The colored bonds indicate the strength of the spin correlator, the endpoints of the bonds represent the position of the correlated sites in the frame of the hole pair, i.e. bond C denotes the spin correlation of next-nearest-neighbours across a hole pair. The data used for the analysis includes up to four holes, but no leg imbalance in the system. \textbf{b}, Quantitative values corresponding to the bonds shown in \textbf{a}. The errorbars denote one s.e.m.}
\label{fig:S4}
\end{center}
\end{figure}

\subsection*{Binding energies in the standard ladder}
For the standard ladder at our parameters we do not find clear signatures of binding from the DMRG calculations (see also below ``\emph{Ground state in a large system}''), i.e. our calculations are consistent with $0\,\Jp$ binding energy. The theoretical binding energies become stronger for larger leg tunnelling \tpl, not only in the standard but also the mixD ladders \cite{bohrdt:2021a}. For the isotropic case $\tpl=\tp$, where we find the largest binding energies in the standard Fermi-Hubbard ladders, our calculations suggest a binding energy of $0.06\,\Jp$ for a ladder system of $L=7$ and interactions strength of $U/\tp = 13.4$ similar to the experimental parameters. This corresponds to an increase in binding in the mixD ladders of about a factor of 13.
Large binding energies almost up to $0.6\,J$, are predicted in the isotropic $t-J$ model, but only in parameter regimes that are not accessible from the pure Fermi-Hubbard model. In these regimes either the required ratio of $t/J$ can not be reached based on Fermi-Hubbard ladders, or the next-nearest-neighbour tunnelling term can not be neglected, as in e.g.~\cite{white:1997a,Chernyshev1998, Schulz1999} (see also above ``\emph{From the Fermi-Hubbard to the $t-J$ model}''). For the isotropic standard Fermi-Hubbard ladder we find, in agreement with~\cite{Karakonstantakis2011}, that binding energies become higher for lower interactions $U/\tp$ but do not exceed $0.25J$. The mixD setting we are employing circumvents these discrepancies and reaches much larger binding energies.

\subsection*{Spin environment around a hole pair}
When a hole pair moves through the system, it locally stretches the antiferromagnetic (AFM) pattern in leg direction, analogously to spin-charge separation in 1D chains~\cite{hilker:2017a}. This should lead to a phase shift of $\pi$ in the AFM pattern across the hole pair. Fig.~\ref{fig:S4} shows the spin correlations $C(\boldsymbol{r}_1,\boldsymbol{r}_2)$ in the reference frame of a hole pair. The negative correlation at distance $d=2$ over the hole pair, and the positive correlation at $d=3$ over the hole pair are reversed in sign compared to an AFM pattern and thus consistent with such a phase jump over the hole pair. This indicates significant mobility of the hole pair along the leg.

\begin{figure}[b]
\begin{center}
\includegraphics{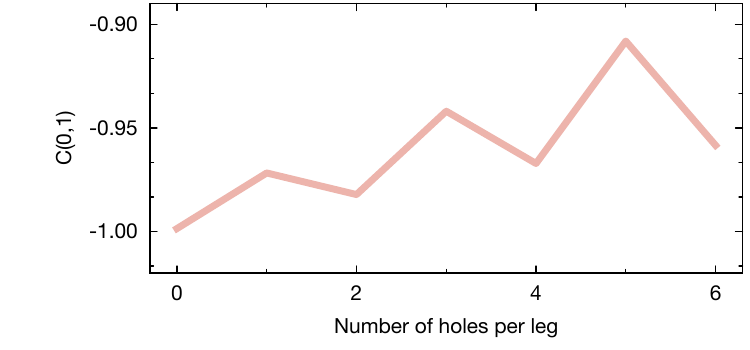}
\caption{
	\textbf{Spin rung correlation vs. number of holes at low temperature.}
	Spin rung correlation $C(0,1)$ in the mixD ladder at a temperature of $0.1 \Jp$ depending on the number of holes in the system calculated using MPS for system length $L=7$ and $t_\parallel = 0.6 J_\perp$. }
\label{fig:S8}
\end{center}
\end{figure}

\subsection*{Even vs. odd number of holes}

In our finite-size system the parity of the total hole number has a significant effect on the properties of the mixD system. For odd hole number, at least one hole has to stay unpaired, because it lacks a partner. This effect can be seen in the staggered behaviour of the singlet strength at low temperature (see Fig.~\ref{fig:S8}). The leftover unpaired hole destroys singlet bonds due to its mobility along the leg direction. The bonds can be restored when an additional hole is added to the system and all holes can pair. We see a similar signature for low hole number in the experimental data (Fig.~\ref{fig:fig2}d).

\begin{figure}[t]
\begin{center}
\includegraphics{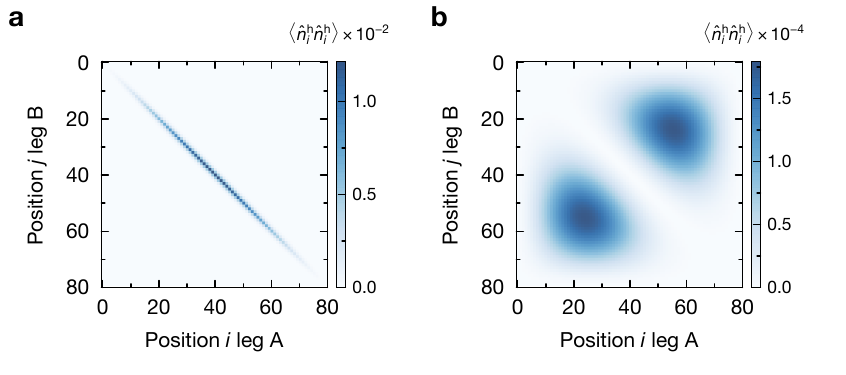}
\caption{
	\textbf{Hole correlation in a larger system.}
	Bare hole correlation $\langle \hat{n}^\mathrm{h}_{i,\mathrm{A}} \hat{n}^\mathrm{h}_{j,\mathrm{B}}\rangle$ for a system of length $L=80$ and $N_{\mathrm{h}}=2$ holes. The data is calculated using DMRG with the same coupling parameters as the experimental data. \textbf{a}, MixD system. Holes are bound in tight pairs, which occupy mostly the central area of the system. \textbf{b}, Standard system. Holes avoid each other and the region towards the edge of the system.}
\label{fig:S10}
\end{center}
\end{figure}

\subsection*{Ground state in a large systems}
Since the experimental system is limited to a length of $L=7$, we compute the ground state of a larger system of $L=80$ containing exactly two holes using DMRG.
We evaluate the bare hole correlator $\langle \hat{n}^\mathrm{h}_{i,\mathrm{A}} \hat{n}^\mathrm{h}_{j,\mathrm{B}}\rangle$ (see Fig.~\ref{fig:S10}) and compare the qualitative signatures to the experimental data.
In the mixD system, we find holes predominantly on the same rung, with a small contribution on neighbouring rungs.
At a distance of three rungs the correlation signal has disappeared, meaning that even in the large system holes are tightly bound.
The pairs furthermore mostly occupy the central area of the system and avoid the region towards the edge, due to its kinetic energy cost. This indicates considerable degree of mobility of the hole pairs.

In the standard system holes avoid both each other and the system edge, to a level where one hole almost exclusively occupies one half of the ladder, and the other one occupies the other half.
In this system size there are still no clear signs of pairing, as no attraction between the holes is visible.

\subsection*{Pair interaction: system length and coupling strength dependence}

\begin{figure}[t]
\begin{center}
\includegraphics{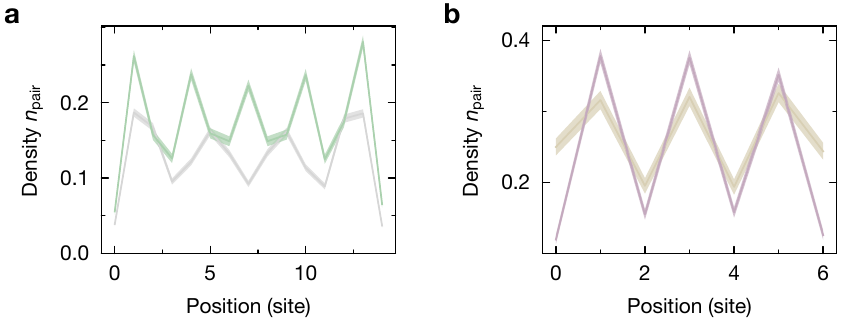}
\caption{
	\textbf{Pair density for different system size and leg coupling.}
	\textbf{a}, Pair density at $\kB T=0.1J_\perp$ for a system length of $L=15$ and $t_\parallel = 0.9 J_\perp$ for eight holes (grey line) and 10 holes (green line) in the system.
	\textbf{b}, Pair density at $\kB T=0.1J_\perp$ for 6 holes in the $L=7$ system for $t_\parallel = 0.6 J_\perp$ (yellow line) and $t_\parallel = 1.2 J_\perp$ (purple line). Both plots are calculated using MPS. Shaded areas denote one s.e.m.}
\label{fig:S6}
\end{center}
\end{figure}

The pair-pair interaction in our MPS calculations shows the same qualitative features as in Fig.~\ref{fig:fig4}b when we vary the system parameters (see Fig.~\ref{fig:S6}).
Increasing the system size to $L=15$ at fixed hole number changes the wavelength of the pair density modulation, but increasing the system size at fixed hole density keeps the wavelength constant (Fig.~\ref{fig:S6}a). 
At hole density $n_\text{h}=1/3$ we find the distance between peaks in rung pair density to be 3 sites, at $n_\text{h}\approx 1/4$ we find a distance of 4 sites, in agreement with the $L=7$ system.
Increasing the leg tunnelling (Fig.~\ref{fig:S6}b) leads to a slightly more pronounced rung pair density modulation, while the position of peaks stays the same. 
In particular the repulsion from the system edge increases. 
Note that, for simplicity, we identify the bound state as pairs of holes on the same rung, although the size of the bound state is expected to grow with increasing leg tunnelling.

\end{document}